\documentclass{article}

    \PassOptionsToPackage{numbers, compress}{natbib}

    \usepackage[preprint]{neurips_2022}

\usepackage[utf8]{inputenc} 
\usepackage[T1]{fontenc}    
\usepackage{hyperref}       
\usepackage{url}            
\usepackage{booktabs}       
\usepackage{amsfonts}       
\usepackage{nicefrac}       
\usepackage{microtype}      
\usepackage{xcolor}         
\usepackage{graphicx}
\usepackage{adjustbox}

\title{Solvent: A Framework for Protein Folding}

\author{%
  Jaemyung Lee \\
  Kakao Brain\\
    \texttt{james.brain@kakaobrain.com} \\
  \And
  Kyeongtak Han\thanks{Kyeongtak Han was an intern at Kakao Brain.}\ \\
  Department of Electrical and Computer Engineering, Inha University \\
  \texttt{han00127@inha.edu} \\
  \And
  Jaehoon Kim \\
  Kakao Brain \\
  \texttt{jack.brain@kakaobrain.com} \\
  \And
  Hasun Yu \\
  Kakao Brain \\
  \texttt{shawn.yu@kakaobrain.com} \\
  \And
  Youhan Lee \\
  Kakao Brain \\
  \texttt{youhan.lee@kakaobrain.com} \\
}

\begin{document}

\maketitle

\begin{abstract}

  Consistency and reliability are crucial for conducting AI research. Many famous research fields, such as object detection, have been compared and validated with solid benchmark frameworks. After AlphaFold2, the protein folding task has entered a new phase, and many methods are proposed based on the component of AlphaFold2. The importance of a unified research framework in protein folding contains implementations and benchmarks to consistently and fairly compare various approaches. To achieve this, we present \textbf{Solvent}, a protein folding framework that supports significant components of state-of-the-art models in the manner of an off-the-shelf interface \textbf{Solvent} contains different models implemented in a unified codebase and supports training and evaluation for defined models on the same dataset. We benchmark well-known algorithms and their components and provide experiments that give helpful insights into the protein structure modeling field. We hope that \textbf{Solvent} will increase the reliability and consistency of proposed models and give efficiency in both speed and costs, resulting in acceleration on protein folding modeling research. The code is available at \href{https://github.com/kakaobrain/solvent}{https://github.com/kakaobrain/solvent}, and the project will continue to be developed.
\end{abstract}

\section{Introduction}


Deep-learning-based protein structure prediction task has become an essential research area after the witnesses of Alphafold2\cite{AlphaFold2021}. Additionally, protein folding methods without multiple sequence alignments(MSA) inputs such as ESMFold\cite{esmfold} and OmegaFold\cite{OmegaFold} have been proposed to model general proteins in high-speed from the removal of generating MSA. Especially, those models have shown promising results on specific applications, including antibody structure and de-novo structure IgFold\cite{igfold} and tFold-Ab\cite{tfold-ab}. Each method has published papers and released inference code with pre-trained models. However, compared to the MSA-dependent method, which has a well-established open-sourced training framework, OpenFold~\cite{ahdritz2022openfold}, the released code of the mentioned MSA-free folding frameworks does not contain training codes, including the input data preparation pipeline. In addition to this, different implementations they used can produce slightly different results for the same components, making clear comparisons difficult. Furthermore, training and evaluating different datasets makes it difficult to compare methods fairly. Reproducing studies in a unified way and comparing them to other methods provides insights to researchers and allows them to propose new methods which are not easily accessible in protein folding.

Because there have been similar difficulties in artificial intelligence fields, it is helpful to study how researchers have overcome them. We bring object detection as a representative example because a well-established framework achieves generalizability and fair comparison for the field. After the development of Faster R-CNN\cite{faster-rcnn}, various R-CNN-based object detection methods have been proposed. The individual method verified its contribution as a publication and source code. However, it was not easy to compare them across studies which is conducted on different codebases. Fortunately, the generalizability and consistency of experiments have been resolved since frameworks like Detectron2\cite{wu2019detectron2} and MMDetection\cite{mmdetection} were proposed. Benchmarking was consistently and user-friendly performed within the same framework for various model variants. The framework provided unified datasets, evaluation metrics, and module implementations. With all other conditions fixed, the effect of the main contribution was clearly and definitely assessed as the frameworks have been used as a de-facto base. In addition,  the framework generalized various models as meta-architecture, which is made up of abstracted components. It reduced the complexity of understanding the object detection pipeline and led to high-quality research, allowing researchers to focus on their ideas, and finally accelerating the field.

The protein folding field needs to mature in the same direction as object detection. Like the emergence of Faster R-CNN in object detection, AlphaFold2 has emerged in protein structure prediction. Based on the modules from Alphafold2, single-sequence-based structure prediction models are being actively explored. To achieve acceleration on the protein folding research as Detectron2 or MMDetection, we present \textbf{Solvent}, the protein folding framework that contains major neural networks, which are the main parts of state-of-the-art models. In the \textbf{Solvent}, several methods are implemented with a unified codebase and represented under meta-architecture. In addition, well-defined datasets are provided for training and validating models. To make \textbf{Solvent} work as a framework, we borrow the pipeline of Detectron2\cite{wu2019detectron2}, which guarantees the consistency and generalizability on \textbf{Solvent}. In the framework, we represent individual methods using the implementation of OpenFold\cite{ahdritz2022openfold}, which is the most reliable and well-known Alphafold2 re-implementation project. We abstract the single sequence-based folding method into three Embedder, Trunk, and Folding module components. We design meta-architecture that one model of ESMFold\cite{esmfold}, OmegaFold\cite{OmegaFold}, and IgFold\cite{igfold} can be called according to the specific types of Embedder, Trunk, and Folding module. Also, the specific algorithm of each component can be selected and combined interchangeably and user-friendly so that new model variants can be easily defined. Furthermore, a researcher can implement a new type of component and combine it with other built-in components in \textbf{Solvent}. For example, a new proposed protein language model can be applied as Embedder and experimented with the existing Trunk and Folding module in \textbf{Solvent}, allowing researchers efficiency in research. In addition to this, \textbf{Solvent} provides built-in support for several train and test datasets to benchmark model performance. Single-chain-based general protein and multi-chain-based antibody datasets are supported with the appropriate metrics. Especially to maximize the training and inference efficiency, we utilized the power of the recently proposed optimization technique, xformers~\cite{xFormers2022}. Also, other optimizations are employed so that the training speed of \textbf{Solvent} is optimized by about 30 \% compared to the original implementations.

To confirm and show how \textbf{Solvent} works, we first experimented with reproducing ESMFold to check the reproducibility of the \textbf{Solvent}. We also experimented with combinations of methods for the components that comprise the meta-architecture to evaluate which Embedder and which Trunk are more effective. Furthermore, we conduct experiments that provide helpful insights for structure prediction studies, such as whether antibody-specific language models can be replaced by general protein language models and how effective Evoformers are. \textbf{Solvent} will be extended to support more algorithms and broader concepts beyond single-sequence protein folding.

\section{Supporting Components}
\label{gen_inst}

\textbf{Solvent} is designed to train and evaluate arbitrary models on arbitrary data, with models and data managed as independent pipelines. This section describes each in more detail.

\subsection{Models}

\textbf{Solvent} supports several different protein folding models, but each model is abstracted into a meta-architecture. The meta-architecture is composed of the following components.

\paragraph{Embedder}

Embedder takes a sequence as input and outputs a sequence embedding with their pair representation computed from pre-trained protein language models (PLM) such as ESM-2. \textbf{Solvent} supports ESM-2\cite{esm-2}, OmegaPLM\cite{OmegaFold}, Antiberty\cite{antiberty} as a built-in embedders.

\paragraph{Trunk}

The trunk is the main building block of structure prediction. It exchanges information between sequence embedding and their pair representation computed by the previous component, Embedder, and includes a recycling embedder that utilizes predicted atom positions from the previous cycle. Evoformer\cite{AlphaFold2021}, GeoformerLite, and IgFoldTrunk\cite{igfold} are built-in supported in the \textbf{Solvent}. Due to GPU memory constraints, we provide GeoformerLite, a simplified version of Geoformer\cite{OmegaFold}. We note that it does not show the full performance of the original Geoformer.

\paragraph{Folding}

The folding module takes single representation and pair representation computed from the previous component, Trunk, and directly predicts the 3D-coordinates of the structure. AlphafoldStructure\cite{AlphaFold2021}, IgfoldStructure\cite{igfold} is supported in the \textbf{Solvent}. 

\paragraph{Heads}

Heads perform task-specific prediction and loss calculation based on features obtained from the Trunk and Folding module(ex. pLDDT, distogram). \textbf{Solvent} includes all the auxiliary heads used in Alphafold2\cite{AlphaFold2021} and IgFold\cite{igfold}. 

All mentioned components are listed in table~\ref{model-abstraction}. Researcher creates model variants easily by changing specific method in the component. For example, a new model can be defined by combining off-the-shelf components, such as the ESM-2 Embedder with the GeoformerLite Trunk, making a new model variant quickly and allowing an accurate comparison between Evoformer and GeoformerLite. In addition to built-in methods, researchers can add new custom methods as components and new models can be defined with built-in modules.

\begin{figure*}[t]
\vskip 0.2in
\begin{center}
\centerline{\includegraphics[width=400pt]{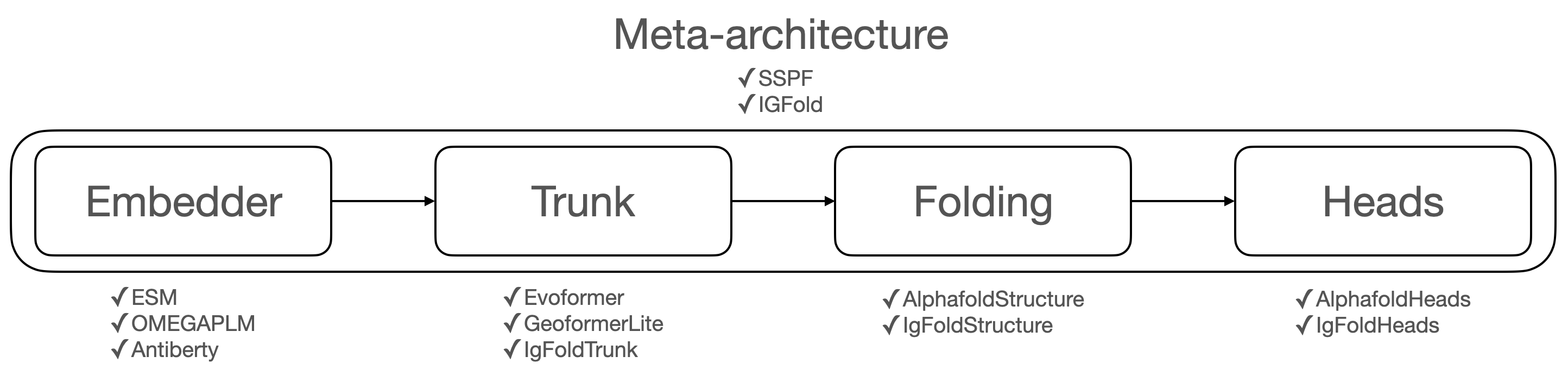}}
\caption{Solvent generalizes model as meta-architecture. The components in meta-architecture is abstracted and their specifics determine folding algorithms.}
\label{meta-arch}
\end{center}
\vskip -0.2in
\end{figure*}

\begin{table}[t]
  \caption{Model abstraction}
  \label{model-abstraction}
  \centering
  \begin{tabular}{l|l|l|l|l}
    \toprule
    \cmidrule(r){1-2}
    Method          & Embedder     & Trunk       & Folding         & Heads \\
    \midrule
    ESMFold         & ESM        & Evoformer   & AlphafoldStructure        & AlphafoldHeads    \\
    OmegaFold-lite  & OmegaPLM     & GeoformerLite   & AlphafoldStructure        & AlphafoldHeads    \\
    IgFold          & Antiberty    & IgFoldTrunk & IgFoldStructure   & IgFoldHeads   \\
    \bottomrule
  \end{tabular}
\end{table}

\subsection{Datasets}

\textbf{Solvent} supports single-chain based general protein dataaset and multi-chain antibody dataset. The datasets described below.

\subsubsection{General protein datasets}

\paragraph{PDB}

The dataset is from Protein Data Bank\cite{pdb} and we use the data before May, 2020 as same as the  ESMFold\cite{esmfold} paper does.

\paragraph{Uniref}

The dataset is basically from Alphafold predicted dataset(afdb)\cite{AlphaFold2021} and we use only the data corresponding to Uniref50. We use the samples with an average pLDDT of 70 or higher.

\paragraph{CAMEO}

The dataset is from CAMEO\cite{cameo} and we use it mainly as evaluation. We use only the data for the 3-months prior to June 25, 2022.

\subsubsection{Antibody datasets}

\paragraph{SAbDab\_20210331}

Antibody dataset based on SAbDab\cite{sabdab} and uses data before March, 2021. Heavy-light chain paired samples and Heavy chain only nanobody samples are included in the dataset.

\paragraph{SAbDab\_igfold}

The selected dataset by IgFold\cite{igfold} paper for the benchmark is a common set used in many papers. We use it mainly as an evaluation.

Except for built-in datasets, custom datasets can be added to the \textbf{Solvent} easily and used to train models with built-in datasets. Any model can be trained and evaluated with well-registered datasets.



\section{Benchmark}
\label{headings}

First, we benchmark the reproduction of ESMFold\cite{esmfold} to check the reproducibility of Solvent. Then we experiment with combinations of different types of Embedder and Trunk for experiencing the convenience of the Solvent and benchmarking of simple model variants. Furthermore, we conduct additional experiments to gain insights to help with protein structure prediction.

\subsection{Experimental settings}

\paragraph{Datasets}

For general protein, we use \verb+pdb+ and \verb+af2_uniref50+ for training models. We use \verb+cameo+ dataset for evaluation. For antibody-specific models, we use \verb+sabdab_20210331+ dataset and \verb+sabdab_igfold+ dataset for training and evaluating models, respectively.

\paragraph{Training Details}

For general proteins, we followed the training scheme of ESMFold\cite{esmfold}. The crop size of the sequence is fixed at 256 for initial training and 384 for finetuning. Since the models we experiment with have low Trunk depth, we do not apply violation loss even when fine-tuning. The batch size is fixed at 128, as in AlphaFold2, and the batch size per GPU depends on the model size.

\paragraph{Evaluation Metrics}

TMscore\cite{tmscore} is used for evaluating general proteins. Sequence alignment is used as a default option. For antibody models, region-specific RMSD is measured using PyRosetta\cite{pyrosetta}.

\subsection{Benchmark existing models}

\subsubsection{Reproducibility of Solvent}

We define ESMFold through Solvent and benchmark the performance for checking reproducibility. Rather than reproducing ESMFold full model with 48 Evoformers, we use Trunk-off models. Different size(35M, 150M, 650M) of ESM-2 is used for experiments. As reported in Table~\ref{reproduce}, some models are reproduced with slightly lower TMscore and some with slightly better performance, but they are reproduced with comparable level to the paper\cite{esmfold}. We train models until finetuning phase only for reproducibility experiments and we train models until initial training for the other experiments.

\begin{table}
  \caption{Comparison ESMFold paper reported TMscore with reproduced ESMFold models by the Solvent. }
  \label{reproduce}
  \centering
  \begin{tabular}{l|ll}
    \toprule
    \cmidrule(r){1-2}
    Method       & Solvent     & Paper reported \\
    \midrule
    ESM-2(35M)   & 0.54        & 0.56    \\
    ESM-2(150M)  & 0.63        & 0.65    \\
    ESM-2(650M)  & 0.71        & 0.70 \\
    \bottomrule
  \end{tabular}
\end{table}

\subsubsection{Combinations of models}

In Solvent, it is easy to combine components of different structure prediction models, making performance comparison between methods possible. Using this convenience, we conduct a combination study of the Embedder and Trunk of two different methods, ESMFold and OmegaFold-lite. Note that the original OmegaFold\cite{OmegaFold} does not share the IPA weight of the structure module, but we experimented with a weight-sharing IPA for solid comparison. As mentioned previous section, we use simplified Geoformer, Geoformer-lite, due to resource constraints. In other words, OmegaFold-lite is not a completely reproduced version of the original OmegaFold. Table~\ref{combinations}  shows the performance of various model variants by permuting components from two different methods. These experiments give researchers a rigorous comparison framework to evaluate which Embedder or Trunk is better.

\begin{table}
  \caption{Simple model vairants by combining component from two different methods.}
  \label{combinations}
  \centering
  \begin{tabular}{l|lll}
    \toprule
    \cmidrule(r){1-2}
    Method          & Embedder          & Trunk             & TMscore \\
    \midrule
    ESMFold         & ESM-2(650M)       & 2 Evoformer       & 0.79 \\
    OmegaFold-lite  & OMEGAPLM(670M)    & 2 GeoformerLite   & 0.75 \\
    Combinations 1  & OMEGAPLM(670M)    & 2 Evoformer       & 0.75 \\
    Combinations 2  & ESM-2(650M)       & 2 GeoformerLite   & 0.78 \\
    \bottomrule
  \end{tabular}
\end{table}

\subsection{Further Study}

The Solvent provides an easy way to conduct various experiments objectively. This allows researchers to make meaningful comparisons between methods. We have conducted some further studies and hope the results give insights for structure prediction research.

\subsubsection{Minimize number of Trunk with trainable language model}


A Trunk module, such as Evoformer, is essential to performance improvement in structure prediction. However, it is computationally expensive, requiring large GPU memory. Therefore, various engineering methods\cite{fastfold} have been proposed to achieve efficiency. A trunk contains an information exchange process between a single representation and a pair representation that contains structural information, and it is repeatedly performed through many blocks (e.g., 48 blocks in ESMFold). Meanwhile, most algorithms such as ESMFold, OmegaFold, and IgFold freeze their PLM layers and let only the Trunk learn the structure from the training data. If we unfreeze the parameters of PLMs and let the structure information be backpropagated into the language model, we can reduce the number of blocks of the Trunk. To prove this, we define four different ESMFold variants based on the type of PLMs and the number of Trunk. In the table~\ref{trainable}, comparing variant1 and variant2 shows the effect of a trainable PLM. Variant1 and variant3 show the effect of the Evoformer. Trainable PLM results in a 7 percent TMscore performance improvement and the use of single Evoformer results in a 10 percent TMscore performance improvement. Adding a single Evoformer onto the trainable PLM model(variant2) has minor performance improvement. We expect the best model as variant4, but a trainable PLM with Evoformer leads to worse performance. All experiments are conducted using ESM-2(35M) as Embedder.

\begin{table}[ht!]
  \caption{Model variant based on Embedder status and the number Trunk. }
  \label{trainable}
  \centering
  \begin{tabular}{l|lll}
    \toprule
    \cmidrule(r){1-2}
    Method         & Embedder Status   & Trunk            & TMscore \\
    \midrule
    variant1       & Freeze            & 0 Evoformer      & 0.53 \\
    variant2       & Trainable         & 0 Evoformer      & 0.60 \\
    variant3       & Freeze            & 1 Evoformer      & 0.63 \\
    variant4       & Trainable         & 1 Evoformer      & 0.61 \\
    \bottomrule
  \end{tabular}
\end{table}

\begin{table}[ht!]
  \caption{More datasets can be registered and used as testsets.}
  \label{testmore}
  \centering
  \begin{tabular}{l|lll}
    \toprule
    \cmidrule(r){1-2}
    Method  & CASP14  & De-novo  &Orphan \\
    \midrule
    variant1  & 0.36  & 0.63  & 0.49 \\
    variant2  & 0.36  & 0.69  & 0.51 \\
    variant3  & 0.40  & 0.74  & 0.51 \\
    variant4  & 0.38  & 0.75  & 0.52 \\
    \bottomrule
  \end{tabular}
\end{table}

\begin{table}[ht!]
  \caption{Antibody model variants based on PLM and their status. }
  \label{abmodel}
  \centering
  \begin{tabular}{l|lll}
    \toprule
    \cmidrule(r){1-2}
    Method         & Embedder       & Embedder Status   & Meta-arch \\
    \midrule
    IgFold(reproduced)       & Antiberty(25M) & Freeze            & IGFold \\
    IgFold-variant1       & ESM-2(35M)     & Freeze            & IGFold \\
    IgFold-variant2       & ESM-2(650M)    & Freeze            & IGFold \\
    IgFold-variant3       & ESM-2(35M)     & Trainable         & IGFold \\
    \bottomrule
  \end{tabular}
\end{table}

\begin{table}[ht!]
  \caption{The performance of various antibody models. }
  \label{abperform}
  \begin{center}
  \begin{adjustbox}{width=360pt}
  \centering
  \begin{tabular}{l|llllllllll}
    \toprule
    \cmidrule(r){1-2}
    Method              & lDDT-C$\alpha$ & OCD & H Fr & H1 & H2 & H3 & L Fr & L1 & L2 & L3 \\
    \midrule
    IgFold(paper)       &       & 3.77 & 0.45 & 0.80 & 0.75 & 2.99 & 0.45 & 0.83 & 0.51 & 1.07 \\
    IgFold(reproduced)  & 0.93  & 3.74 & 0.57 & 0.92 & 0.80 & 3.09 & 0.67 & 1.12 & 0.55 & 1.15 \\
    IgFold-variant1     & 0.92  & 3.76 & 0.62 & 0.87 & 0.94 & 3.06 & 0.49 & 0.90 & 0.51 & 1.15 \\
    IgFold-variant2     & 0.93  & 3.77 & 0.48 & 0.91 & 0.94 & 3.20 & 0.48 & 0.94 & 0.49 & 1.13 \\
    IgFold-variant3     & 0.93  & 3.88 & 0.51 & 0.89 & 0.85 & 3.14 & 0.51 & 1.00 & 0.50 & 1.10 \\
    \bottomrule
  \end{tabular}
  \end{adjustbox}
  \end{center}
\end{table}

\subsubsection{Use general protein language model on antibody structure predictions}

Various large-scale language models exist for general proteins, such as ESM-2(up to 15B) and OmegaPLM(670M). However, antibody-specific models are represented by Antiberty(25M), which is small compared to general proteins, and the size of the dataset to train them is also small compared to ESM-2 and OmegaPLM. Currently, the general protein language model is publicly available and easy to use. It is worth investigating whether Antiberty, an antibody-specific language model, is still particularly unique for antibody structure prediction compared to general PLM. The details of antibody model variants are listed in Table~\ref{abmodel}. The performance of the listed model is reported in Table~\ref{abperform}.


From the comparison between IgFold(reproduced) and Igfold-variant1, the performance difference between using Antiberty and ESM-2(35M) is not very noticeable. The Antiberty model performs better on some CDRs but does not significantly. In fact, using a large general protein language model(IgFold-variant2) model seems more effective than using an antibody-specific language model. Using general language model with trainable parameters(IgFold-variant3) does not show performance improvement on most CDRs. 


\subsection{Custom datasets can be added and evaluated on different models}

In \textbf{Solvent}, any datasets can be registered and used for training and evaluating models. As an example, we register CASP14 datasets, de-novo proteins, and orphan proteins. In the case of CASP14, we used 33 publically released samples. T1044 is not included due to memory constraints. In the case of de-novo and orphan proteins, we referred to the target lists provided at RGN2\cite{RGN2} repository and used samples that were released after May 2020. These samples might be used when training ESM-2, which causes high performance for de novo proteins. All the samples are listed in the appendix. The model variants listed in Table~\ref{trainable} can be evaluated on these three different datasets(Table~\ref{testmore}). 

\section{Conclusion}

To support a consistent and easy-to-use research framework, we propose \textbf{Solvent} for protein folding research. We hope that efficient and rigorous experiments on top of the \textbf{Solvent} will further prove the strengths and weaknesses of each algorithm and finally accelerate structural prediction research. Currently, \textbf{Solvent} focuses on MSA-free protein structure prediction models.  We will extend \textbf{Solvent} to a more general way that takes MSA and template as input and support more validation data such as orphan and de-novo protein.


\begin{ack}
We acknowledge the contributions of the Language Model Engineering Team at Kakao Brain, who have optimized Solvent. These optimizations make Solvent efficient in training speed and memory, so researchers can easily tap larger models. Their support has been essential in achieving the outcomes presented in this work.

\end{ack}

\bibliographystyle{unsrt}
\bibliography{ref}

\begin{thebibliography}{10}

\bibitem{AlphaFold2021}
John Jumper, Richard Evans, Alexander Pritzel, Tim Green, Michael Figurnov,
  Olaf Ronneberger, Kathryn Tunyasuvunakool, Russ Bates, Augustin
  {\v{Z}}{\'\i}dek, Anna Potapenko, Alex Bridgland, Clemens Meyer, Simon A~A
  Kohl, Andrew~J Ballard, Andrew Cowie, Bernardino Romera-Paredes, Stanislav
  Nikolov, Rishub Jain, Jonas Adler, Trevor Back, Stig Petersen, David Reiman,
  Ellen Clancy, Michal Zielinski, Martin Steinegger, Michalina Pacholska, Tamas
  Berghammer, Sebastian Bodenstein, David Silver, Oriol Vinyals, Andrew~W
  Senior, Koray Kavukcuoglu, Pushmeet Kohli, and Demis Hassabis.
\newblock Highly accurate protein structure prediction with {AlphaFold}.
\newblock {\em Nature}, 596(7873):583--589, 2021.

\bibitem{esmfold}
Zeming Lin, Halil Akin, Roshan Rao, Brian Hie, Zhongkai Zhu, Wenting Lu, Nikita
  Smetanin, Robert Verkuil, Ori Kabeli, Yaniv Shmueli, Allan dos Santos~Costa,
  Maryam Fazel-Zarandi, Tom Sercu, Salvatore Candido, and Alexander Rives.
\newblock Evolutionary-scale prediction of atomic level protein structure with
  a language model.
\newblock {\em bioRxiv}, 2022.

\bibitem{OmegaFold}
Ruidong Wu, Fan Ding, Rui Wang, Rui Shen, Xiwen Zhang, Shitong Luo, Chenpeng
  Su, Zuofan Wu, Qi~Xie, Bonnie Berger, Jianzhu Ma, and Jian Peng.
\newblock High-resolution de novo structure prediction from primary sequence.
\newblock {\em bioRxiv}, 2022.

\bibitem{igfold}
Jeffrey~A Ruffolo, Lee-Shin Chu, Sai~Pooja Mahajan, and Jeffrey~J Gray.
\newblock Fast, accurate antibody structure prediction from deep learning on
  massive set of natural antibodies.
\newblock {\em Nature communications}, 14(1):2389, 2023.

\bibitem{tfold-ab}
Jiaxiang Wu, Fandi Wu, Biaobin Jiang, Wei Liu, and Peilin Zhao.
\newblock tfold-ab: Fast and accurate antibody structure prediction without
  sequence homologs.
\newblock {\em bioRxiv}, 2022.

\bibitem{ahdritz2022openfold}
Gustaf Ahdritz, Nazim Bouatta, Sachin Kadyan, Qinghui Xia, William Gerecke,
  Timothy~J O’Donnell, Daniel Berenberg, Ian Fisk, Niccol{\`o} Zanichelli,
  Bo~Zhang, et~al.
\newblock Openfold: Retraining alphafold2 yields new insights into its learning
  mechanisms and capacity for generalization.
\newblock {\em bioRxiv}, pages 2022--11, 2022.

\bibitem{faster-rcnn}
Shaoqing Ren, Kaiming He, Ross Girshick, and Jian Sun.
\newblock Faster r-cnn: Towards real-time object detection with region proposal
  networks.
\newblock In C.~Cortes, N.~Lawrence, D.~Lee, M.~Sugiyama, and R.~Garnett,
  editors, {\em Advances in Neural Information Processing Systems}, volume~28.
  Curran Associates, Inc., 2015.

\bibitem{wu2019detectron2}
Yuxin Wu, Alexander Kirillov, Francisco Massa, Wan-Yen Lo, and Ross Girshick.
\newblock Detectron2.
\newblock \url{https://github.com/facebookresearch/detectron2}, 2019.

\bibitem{mmdetection}
Kai Chen, Jiaqi Wang, Jiangmiao Pang, Yuhang Cao, Yu~Xiong, Xiaoxiao Li,
  Shuyang Sun, Wansen Feng, Ziwei Liu, Jiarui Xu, Zheng Zhang, Dazhi Cheng,
  Chenchen Zhu, Tianheng Cheng, Qijie Zhao, Buyu Li, Xin Lu, Rui Zhu, Yue Wu,
  Jifeng Dai, Jingdong Wang, Jianping Shi, Wanli Ouyang, Chen~Change Loy, and
  Dahua Lin.
\newblock Mmdetection: Open mmlab detection toolbox and benchmark.
\newblock {\em CoRR}, abs/1906.07155, 2019.

\bibitem{xFormers2022}
Benjamin Lefaudeux, Francisco Massa, Diana Liskovich, Wenhan Xiong, Vittorio
  Caggiano, Sean Naren, Min Xu, Jieru Hu, Marta Tintore, Susan Zhang, Patrick
  Labatut, and Daniel Haziza.
\newblock xformers: A modular and hackable transformer modelling library.
\newblock \url{https://github.com/facebookresearch/xformers}, 2022.

\bibitem{esm-2}
Zeming Lin, Halil Akin, Roshan Rao, Brian Hie, Zhongkai Zhu, Wenting Lu, Nikita
  Smetanin, Allan dos Santos~Costa, Maryam Fazel-Zarandi, Tom Sercu, Sal
  Candido, et~al.
\newblock Language models of protein sequences at the scale of evolution enable
  accurate structure prediction.
\newblock {\em bioRxiv}, 2022.

\bibitem{antiberty}
Jeffrey~A Ruffolo, Jeffrey~J Gray, and Jeremias Sulam.
\newblock Deciphering antibody affinity maturation with language models and
  weakly supervised learning.
\newblock {\em arXiv}, 2021.

\bibitem{pdb}
Helen~M. Berman, John Westbrook, Zukang Feng, Gary Gilliland, T.~N. Bhat, Helge
  Weissig, Ilya~N. Shindyalov, and Philip~E. Bourne.
\newblock {The Protein Data Bank}.
\newblock {\em Nucleic Acids Research}, 28(1):235--242, 01 2000.

\bibitem{cameo}
J{"u}rgen Haas, Alessandro Barbato, Dario Behringer, Gabriel Studer, Steven
  Roth, Martino Bertoni, Khaled Mostaguir, Rafal Gumienny, and Torsten Schwede.
\newblock Continuous automated model evaluation (cameo) complementing the
  critical assessment of structure prediction in casp12.
\newblock {\em Proteins}, 86(S1):387--398, 2018.

\bibitem{sabdab}
James Dunbar, Konrad Krawczyk, Jinwoo Leem, Terry Baker, Angelika Fuchs, Guy
  Georges, Jiye Shi, and Charlotte~M. Deane.
\newblock {SAbDab: the structural antibody database}.
\newblock {\em Nucleic Acids Research}, 42(D1):D1140--D1146, 11 2013.

\bibitem{tmscore}
Y.~Zhang and J.~Skolnick.
\newblock Scoring function for automated assessment of protein structure
  template quality.
\newblock {\em Proteins}, 57(4):702--710, 2004.

\bibitem{pyrosetta}
Sidhartha Chaudhury, Sergey Lyskov, and Jeffrey~J. Gray.
\newblock Pyrosetta: a script-based interface for implementing molecular
  modeling algorithms using rosetta.
\newblock {\em Bioinformatics}, 26(5):689--691, 2010.

\bibitem{fastfold}
Shenggan Cheng, Ruidong Wu, Zhongming Yu, Binrui Li, Xiwen Zhang, Jian Peng,
  and Yang You.
\newblock Fastfold: Reducing alphafold training time from 11 days to 67 hours,
  2022.

\bibitem{RGN2}
Ratul Chowdhury, Nazim Bouatta, Surojit Biswas, Charlotte Rochereau, George~M.
  Church, Peter~K. Sorger, and Mohammed AlQuraishi.
\newblock Single-sequence protein structure prediction using language models
  from deep learning.
\newblock {\em bioRxiv}, 2021.

\end{thebibliography}

\appendix

\section{Appendix}

\subsection{List of samples used in Table~\ref{testmore}}

\subsubsection{CASP14}

We used 33 publicaly released samples. T1044 is not included. \textit{T1024, T1025, T1026, T1027, T1029, T1030, T1031, T1032, T1033, T1035, T1036s1, T1037, T1038, T1039, T1040, T1041, T1042, T1043, T1046s1, T1046s2, T1049, T1050, T1054, T1056, T1064, T1067, T1073, T1074, T1079, T1080, T1082, T1090, T1099}.

\subsubsection{De novo proteins}

From the RGN2 repository target lists, we used 111 samples released after May, 2020. \textit{7BQS\_A, 6X1K\_A, 6W6X\_A, 7A4Y\_E, 7BQR\_A, 7RX5\_A, 7BQC\_A, 7BQQ\_A, 7DNS\_A, 7BQB\_A, 6YPI\_A, 6Z35\_A, 7BQE\_A, 6ZOF\_D, 7CZ0\_E, 7BPN\_A, 7BQD\_A, 7CX4\_N, 7BPM\_A, 6WXO\_B, 6XT4\_A, 6XEH\_A, 7BPL\_A, 7BQN\_A, 6XSS\_A, 6WMK\_A, 7AWY\_A, 7AVA\_A, 7A4D\_E, 7BO9\_C, 7DI0\_A, 6YB2\_A, 7BO8\_B, 6YB1\_A, 6VG7\_A, 7BIM\_A, 7BEY\_A, 7AWZ\_A, 6YB0\_A, 7A8S\_A, 7BQM\_A, 6RLH\_A, 6WXP\_A, 6W9Y\_A, 6UIB\_C, 6RLI\_A, 6U1S\_A, 7RGR\_B, 6WVS\_A, 7BPP\_A, 6W9Z\_D, 7ARR\_A, 6ZL1\_C, 7AX0\_A, 7ARS\_A, 6XI6\_A, 6ZT1\_A, 6WRV\_D, 7AX2\_A, 6Y7N\_A, 6VFL\_B, 6WRW\_C, 6VGA\_A, 6VFK\_B, 6Z3X\_B, 7BNT\_B, 7CUV\_A, 7N8J\_B, 6VFJ\_B, 7KBQ\_A, 7CD8\_A, 6YAZ\_E, 6VFI\_B, 6X8N\_A, 6VGB\_A, 6VFH\_A, 6W70\_A, 7BAU\_E, 6ZIE\_A, 6Z0L\_A, 6REO\_A, 6ZV9\_B, 6W90\_A, 7BAT\_A, 7JH6\_A, 6Z0M\_A, 7A50\_B, 7JH5\_A, 7BAW\_A, 6YWC\_C, 6WA0\_B, 7DDR\_A, 6Z2I\_A, 7BWW\_A, 7A48\_B, 7BAV\_C, 6WRX\_C, 6V67\_B, 6X9Z\_A, 6YWD\_C, 6XR1\_A, 6XNS\_B, 6VL5\_D, 6VEH\_A, 7A1T\_C, 6XR2\_E, 6VL6\_A, 7BAS\_B, 7CBC\_A, 6XH5\_B, 7BOA\_C}

\subsubsection{Orphan proteins} 
From the RGN2 repository target lists, we used 6 samples released after May, 2020. \textit{6M64\_F, 6YNS\_O, 7AL0\_A, 7DGU\_A, 7KEI\_C, 7OSC\_A}.

\end{document}